\documentclass[a4paper,twoside]{article}
\newcommand{\affil}[1]{$^{\rm #1}$}
\usepackage{graphicx}
\date{} 
\title{\large\bf\flushleft What is a Galaxy? Cast your vote here...}
\author{\parbox{\textwidth}{\flushleft
\vspace{-0.5cm}
{\it Duncan A. Forbes\affil{A,C} and Pavel Kroupa\affil{B}
}\\
\vspace{0.4cm}
{\small \affil{A}\,Centre for Astrophysics \& Supercomputing, Swinburne University, Hawthorn VIC 3122, Australia}\\
{\small \affil{B}\,Argelander Institute for Astronomy, University of Bonn, Auf dem Hugel 71, D-53121 Bonn, Germany}\\
{\small \affil{C}\,Email: dforbes@swin.edu.au}}}
\begin{document}
\twocolumn[
\begin{changemargin}{.8cm}{.5cm}
\begin{minipage}{.9\textwidth}
\vspace{-1cm}
\maketitle
%
%
\small{\bf Abstract: 
Although originally classified as galaxies, Ultra Compact Dwarfs
(UCDs) share many properties in common with globular star
clusters. The debate on the origin and nature of UCDs, and the
recently discovered ultra-faint dwarf spheroidal (dSph) galaxies
which contain very few stars,
has motivated us to ask the question `what is a galaxy?'
Our aim here is to promote further discussion of how to define a
galaxy and, in particular, what separates it from a star cluster. 
Like most previous definitions, we adopt the requirement of 
a gravitationally bound stellar system as a minimum. In order to
separate a dwarf galaxy from a globular cluster, we discuss other
possible requirements, such as a minimum size, a long two-body relaxation
time, a satellite system, the presence of complex stellar
populations and non-baryonic dark matter. We 
briefly mention the implications of each of these definitions if they are
adopted. 
Some special cases of objects with an ambiguous nature are also
discussed. 
Finally, we give our favoured criteria, and 
in the spirit of a `collective wisdom',  
invite readers to vote on their preferred definition
of a galaxy via a dedicated website. 
}

\medskip{\bf Keywords:} galaxies: dwarf -- galaxies: fundamental
parameters -- galaxies: star clusters -- galaxies: general

\medskip
\medskip
\end{minipage}
\end{changemargin}
]
\small

\section{Introduction}

Astronomers like to classify things. That classification may
initially be
based on appearance to the human eye (e.g. Hubble 1926), but to make
progress this taxonomy may need to have some basis in the
underlying nature/physics of the objects being examined. With
this mind, astronomers need a working definition so as to divide objects
into different categories and to explore the interesting
transition cases that might share
common properties. Hopefully this results in 
additional insight into the physical
processes that are operating. 

Perhaps the most famous recent case of classification in 
astronomy is the International
Astronomical Union's definition of a planet and its separation on
small scales from minor bodies in the solar system. This was
partly motivated by the recent discoveries of several planet-like objects
that challenged the previous loose definition of a planet. After 
2 years of preparation by an IAU working group and 2 weeks of debate at an IAU
General Assembly in Prague, the IAU presented its new definition for a
planet. The criteria included a clause that a planet should
dominate its local environment, which Pluto did not, and hence it
was officially removed from its long standing status as a planet
(see also Soter et al. 2006). 
This decision was not uniformly welcomed, especially among the
general public.

There is no widely-accepted standard definition for a galaxy.  
In this paper we discuss the issue of small scale stellar systems,
in particular dwarf galaxies and what separates them from star
clusters. A working definition for a dwarf galaxy was suggested
in 1994 by Tammann, i.e. those galaxies fainter than M$_B$ $=$
--16 
and more extended than
globular clusters. Since that time   
Ultra Compact Dwarf (UCD) objects (Hilker et al. 1999; later called galaxies by
Drinkwater et al. 2000 and 
Phillipps et al. 2001) and ultra-faint dwarf spheroidal (dSph)
galaxies around the Milky Way have been discovered. 
UCDs (also called Intermediate Mass Objects and Dwarf
Galaxy Transition Objects) have properties intermediate between
those  traditionally recognised as galaxies and globular star clusters. 
Whereas some of the ultra-faint dSph galaxies contain so few stars that
they can be fainter than a single bright star  
and contain less stellar mass than some globular clusters (e.g. Belokrov et
al. 2007).  

\section{Ultra Compact Dwarfs: star clusters or galaxies?}

UCDs have sizes, luminosities and masses that are intermediate
between those traditionally classified as globular clusters and
dwarf galaxies (e.g. Dabringhausen et al. 2008; 
Forbes et al. 2008;  Taylor et
al. 2010). Their properties and relationship with `normal' globular
clusters have been reviewed recently by Hilker (2009). 
Although UCDs have similar luminosities and {\it
stellar} masses to dwarf spheroidal galaxies, they are much 
more compact. They have been shown to contain predominately old
aged stars and to be pressure-supported (Chilingarian et al. 2010).
Unlike the lower mass
globular clusters (with a near constant half-light radius of r$_h$
$\approx$ 3 pc),
UCDs reveal a near linear size-luminosity trend (Dabringhausen et al. 2008; 
Forbes et al. 2008;  Taylor et al. 2010).\\

There is no universally-accepted definition of a UCD,
however parameters commonly adopted are:\\

\noindent
$\bullet$ 10 $\le$ r$_h$/pc $\le$ 40\\

\noindent
$\bullet$ --10.5 $\ge$ M$_V$ $\ge$ --14\\

\noindent
$\bullet$ 2 $\times$ 10$^6$ $\le$ M/M$_{\odot}$ $\le$ 2 $\times$ 10$^8$. \\ 

Some workers also apply an ellipticity criterion to ensure near roundness 
(e.g. Madrid et al. 2010). 
We also note that UCD-like objects with half-light sizes up to 100 pc
have been reported as confirmed members of the Virgo, Fornax and Coma clusters
(Evstigneeva et al. 2007; Chiboucas et al. 2010.)

The issue of whether UCDs contain dark matter, or not, is still
subject to debate as the mass-to-light ratios are slightly higher
than expected for a standard IMF with current stellar
population models (Baumgardt \& Mieske 2008; Chilingarian et
al. 2010). 
However, it would only take a small difference
in the IMF or additional cluster evolution physics for the
inferred mass-to-light ratios to be consistent with a purely stellar
system devoid of dark matter (see Dabringhausen et al. 2009 
for further discussion
of this issue). 

Formation scenarios for UCDs include a galaxy origin, i.e. as the
remnant nucleus of a stripped dwarf galaxy (Bekki et al. 2001) or
as the rare surviving relic of a dwarf galaxy formed in the early
Universe (Drinkwater et al. 2004). Star cluster origins include
the merger of several smaller star clusters (Fellhauer \& Kroupa
2002) or that they are simply the extension of the globular
cluster sequence to higher masses (e.g. Mieske et
al. 2004). Multiple origins are also possible (Norris \&
Kannappan 2010; Da Rocha et al. 2010; Chilingarian et al. 2010). 
The
observation that UCDs are consistent with the GC luminosity (mass)
function and follow a similar spatial distribution to GCs around a
host galaxy would argue that {\it most} UCDs are effectively
massive star clusters. So although some interesting exceptions
may exist, we favour the view that UCDs today are dark matter free star
clusters. Whether they should also be called `galaxies' is
discussed below.



\section{The Definition of a Galaxy}

Descriptive definitions of a galaxy are numerous. Below are 
three examples selected from popular websites:\\

\noindent
{\it A galaxy is a massive, gravitationally bound system that
consists of stars 
and stellar remnants, an interstellar medium of gas and dust, 
and an important but poorly understood component 
tentatively dubbed dark matter.}\\
(http://en.wikipedia.org/wiki/Galaxy).\\

\noindent
{\it Any of the numerous large groups of stars and other matter that
exist in space as independent systems.}\\ 
(http://www.oed.com/)\\

\noindent
{\it A galaxy is a gravitationally bound entity, typically consisting
of dark matter, gas, dust and stars.} \\
(http://astronomy.swin.edu.au/cms/astro/cosmos/).\\

Most popular definitions 
require that a galaxy consists of
matter that is gravitationally self-contained or bound. This
matter could take different forms, but the presence of stars
is generally required. 
Taking this as the starting point for the definition of a
galaxy, we require that a galaxy is:\\

\noindent 
{\bf I. Gravitationally bound}\\
A fundamental criterion to be a galaxy is that the
matter must be gravitationally bound (i.e. have a negative binding energy) 
within its own potential well. Matter
that is unbound may
include material stripped away by the action of a tidal encounter
or `evaporated' away if it exceeds the escape velocity of the system.

If being gravitationally bound is a requirement to
be a galaxy, then collections of `tidal material' are not galaxies.\\

\noindent
{\bf II. Contains stars}\\
An additional key requirement is that a galaxy be a stellar system,
i.e. include the presence of some stars. In the case of recently
discovered ultra-faint dwarf spheroidal galaxies, the number of stars
inferred can be as low as a few hundred. 

It is possible, and indeed predicted by some simulations
(e.g. Verde, Oh \& Jimenez 2002), that `dark
galaxies' exist, i.e. dark matter halos that contain cold gas which
has for some reason failed to form any stars. In general 21~cm
radio searches
for such objects indicate that they do not exist in large
numbers, if at all (Kilborn et al. 2005; Doyle et al. 2005).

If the presence of stars is a requirement to
be a galaxy, then gas-rich, star-free 
`dark galaxies' are not galaxies. \\

These two criteria taken together would exclude tidal material
and `dark galaxies', but would however include star clusters,
such as globular clusters and UCDs, in the definition of a galaxy. 
Additional criteria are probably required. A few suggestions and
their implications are listed below. \\

\noindent
{\bf $\bullet$ Two-body relaxation time $\ge$ H$^{-1}_{0}$} \\
When a stellar system is in a stable dynamical state, the 
orbits of the stars are determined by the mean gravity of the
system rather than localised encounters between individual stars. 
In other words, galaxies are long-lasting 
systems with smooth gravitational potentials 
that can be modelled over time by the collisionless Boltzmann
equation (Kroupa 2008).
This can be quantified by calculating the 
two-body relaxation time (Binney \& Tremaine 1987; Kroupa 1998), i.e.\\
$t_{\rm rel} \approx {0.2 \over \sqrt{G}}\,{1\over m_{\rm av}}\, 
{M^{1\over2} \over {\rm ln}M}\,
r^{3\over2}$, \\
where $G=0.0045 {\rm pc}^3 M_\odot^{-1} {\rm
Myr}^{-2}$, 
$M$ and $r$ are the mass (in M$_{\odot}$) and 
characteristic radius (in pc) of the system 
and $m_{\rm av}$ is the average stellar mass (typically
$0.5\,M_\odot$ for old stellar systems) with the relaxation time
given in Myrs. 
Systems with a relaxation time longer than the age of the
Universe would include UCDs ($M>10^6\,M_\odot$ and $r>10\,$pc) 
and tidal dwarf galaxies 
($M>10^4\,M_\odot$ and $r>100\,$pc) 
but not star clusters traditionally
classified as globular clusters ($M<10^6\,M_\odot$ and $r
\approx 3\,$pc) 

If having a relaxation time longer than the Hubble time is a requirement to
be a galaxy, then `globular clusters' are not galaxies but 
`ultra compact dwarfs' and `tidal dwarf galaxies' are galaxies. \\

\noindent
{\bf $\bullet$ Half-light radius  $\ge$ 100 pc} \\
The half-light size, or effective radius, is a useful measure of
the extent of a stellar system. As mentioned above, UCDs
generally have sizes up to 40 pc.
The recently discovered ultra-faint dwarf
spheroidals have sizes as small as 100 pc. 
Thus there appears to be a zone-of-avoidance, which can
not be entirely due to selection effects, of sizes 40 $<$ r$_h$
$<$ 100 pc for which objects are very rare (Gilmore et al. 2007;
Belokurov et al. 2007). 
Gilmore et al. have
argued that the few objects within this zone-of-avoidance
are special cases which are probably not in equilibrium but in
the throes of disruption. 
However the zone-of-avoidance is rapidly being filled, 
with several UCD-like objects have measured half-light sizes
as large as 100 pc.

If having a half-light size greater than $\sim$100 pc is a requirement to
be a galaxy, then `ultra compact dwarfs' (and globular clusters) 
are not galaxies. \\

\noindent 
{\bf $\bullet$ Presence of complex stellar populations}\\
In a sufficiently deep potential well, some gas left over from
the first episode of star formation will remain. This gas, and
more enriched gas from stellar mass loss and supernovae, may be
available for a second episode of star formation. Thus complex
stellar populations of different abundances and ages will be
present in substantial stellar systems. This is in contrast 
to the single stellar populations found in most star
clusters. However recent observational data has revealed clear
evidence for multiple stellar populations in the more massive
Milky Way globular clusters (Piotto 2009). A possible explanation for this is
self-enrichment within a larger proto-cluster gas cloud 
(Parmentier 2004; 
Strader \& Smith 2008; Bailin \& Harris 2009). This self-enrichment
process becomes apparent at masses around one million
solar masses.

If the presence of complex stellar populations is a requirement to
be a galaxy, then massive globular clusters (and probably `ultra compact
dwarfs') are galaxies. \\

\noindent
{\bf $\bullet$ Presence of non-baryonic dark matter}\\
Our standard paradigm of galaxy formation is that every galaxy
formed in a massive dark matter halo (White \& Rees 1978). 
Thus the presence of dark matter is seen by many as a key requirement
to be classified as a galaxy (e.g. Gilmore et al. 2007). 
It is unfortunately a difficult
property to measure empirically for dwarf galaxies, usually
relying on measurements of the velocity dispersion. So although
high dark matter fractions have been inferred for Local Group dSph
galaxies (assuming the velocity dispersion is a valid diagnostic
for these systems), dE galaxies indicate very little dark matter
within the half-light radius (Toloba et al. 2010; Forbes et
al. 2010) and perhaps to $\sim$10 times the half-light
radius in the case of NGC 147 and NGC 185, depending on the choice
of IMF and stellar population model (Geha et al. 2010). 
An alternative explanation to the measured high velocity dispersions is
that non-Newtonian dynamics are operating
(Brada \& Milgrom 2000; McGaugh \& Wolf 2010).

We note that tidal dwarf galaxies, that form out of the
collapse of disk material in a tidal tail after a merger, are not
expected to contain much dark matter (Barnes \& Hernquist 1992;
Gentile et al. 2007). Thus if any of the dSph galaxies
which surround the Milky Way in a disk of satellites (Metz et
al. 2009; Kroupa et al. 2010) have a tidal dwarf origin they
would not be expected to have a high dark matter content (and the observed
velocity dispersions used to infer the presence of dark matter
would be an invalid diagnostic). 

The dark matter 
galaxy formation scenario may extend down to GCs which have been suggested 
to form in (mini) dark matter
halos (Bromm \& Clarke 2002; Mashchenko \& Sills 2005; Saitoh et
al. 2006; Griffen et al. 2010). However, they must have lost this
dark matter as none as been detected to date in Milky Way GCs 
(Moore 1996; Baumgardt et al. 2009; Lane et
al. 2010; Conroy, Loeb \& Spergel 2010).

If the presence of a massive dark matter halo is a requirement to
be a galaxy, then probably `tidal dwarf galaxies', `ultra compact
dwarfs', and possibly some Milky Way `dwarf spheroidal galaxies'
and `dwarf elliptical galaxies' are not galaxies. \\

\noindent
{\bf $\bullet$ Hosts a satellite stellar system}\\
Evidence that a galaxy dominates its environment could come from
the presence of smaller satellite stellar systems, such as dwarf
galaxies (for large galaxies) or globular clusters. All known
large galaxies possess a system of globular clusters, however
some dwarf galaxies do not host any globular clusters
(e.g. Forbes 2005). For example, 
in the Local Group, the dwarf galaxy WLM has a single globular
cluster but the galaxies Aquarius, Tucana and the recently discovered
ultra-faint dwarf satellites of the Milky Way appear to have none.

If the presence of a globular cluster system is a requirement to
be a galaxy, then `ultra compact
dwarfs' and some of the smallest `dwarf galaxies' are not galaxies. \\

Of course we don't live in a static Universe, and an object could
evolve from a galaxy into a star cluster (or viz versa). For
example, it has been suggested that globular clusters may sink to
the centre of a galaxy via dynamical friction forming a galaxy
nucleus. If that galaxy then loses its outer stars from tidal
stripping, leaving only the remnant nucleus it may be classified
as a UCD or a globular cluster. Passive evolution or interactions
(mergers, tidal stripping etc) can change the nature of an object
over time. The criteria listed above apply to objects today and
not their past or future state. 

Below we briefly mention some
special cases of stellar systems which challenge attempts to
define a galaxy.

\section{Special Cases}

\noindent
$\bullet$ {\bf Omega Cen and G1}\\
Omega Cen has traditionally been known as the most massive
globular cluster in the Milky Way system. However, the presence
of multiple stellar populations, its large size, elongation,
Helium abundance and
retrograde orbit have led many to suggest it is actually the
remnant nucleus of a disrupted dwarf galaxy (Freeman 1993; Bekki
\& Freeman 2003). It may therefore represent a (low-mass) example
of a UCD. 
Similar arguments have been made for the
globular cluster G1 in M31 (Meylan et al. 2001; Bekki \& Chiba
2004). Otherwise, both Omega Cen and G1 are consistent with the
general scaling properties of massive globular clusters (e.g. Forbes
et al. 2008). \\

\noindent
$\bullet$ {\bf Willman~1, Segue~1, Segue~2 and Bootes II}\\
Willman~1 (Willman et al. 2005), Segue~1 (Belokurov et al. 2007),
Segue~2 (Belokurov et al. 2009) 
and Bootes II (Walsh, Jerjen \& Willman 2007) are all low surface
brightness objects discovered recently in deep surveys. They 
have low luminosities of M$_V$ $\sim$ --2 (stellar masses of a
few hundred solar masses) and half light sizes of
r$_h$ $\sim$ 30 pc. Such values place them at the extreme of the
globular cluster distribution and with relaxation timescales much
shorter than the age of the Universe. 
In the case of Segue~1, Geha et al.
(2009) suggested it is a galaxy with a mass-to-light ratio of $\sim$1200 on the
basis of a measured velocity dispersion of 4.2 $\pm$ 1.2
km/s. Subsequently,
Niederste-Ostholt et al. (2009) found that this velocity
dispersion may be inflated by nearby Sagittarius dwarf galaxy stars, and
favoured a globular cluster status for Segue~1. Most recently,
Simon et al. (2010) have reiterated that Seque~1 is a dark matter
dominated dwarf galaxy. They conclude that stars from 
the Sagittarius dwarf do not unduely
affect their results and that {\it ...the metallicities of stars in
Segue~1 provide compelling evidence that, irrespective of its
current dynamical state, Segue~1 was once a dwarf galaxy.} \\

\noindent
$\bullet$ {\bf Coma Berenices}\\
Coma Berenices was discovered by Belokurov et al. (2007). Deep
imaging by Munoz, Geha and Willam (2010) indicates 
a half-light size of 
r$_h$ = 74 pc and an ellipticity of 0.36. 
The V band luminosity was determined to be M$_V$ =
--3.8. Thus it has a similar luminosity to Willman~1, Segue~1, 2 and
Bootes II but is significantly larger in size. Its size places it
within the half-light zone-of-avoidance between the locus of
globular clusters/UCDs and dwarf galaxies. However there is no obvious
sign of tidal stripping to faint surface brightness levels. 
Simon \& Geha (2007) derive a metallicity [Fe/H] = --2 with zero
dispersion. However, Kirby et al. (2008) quote a mean metallicity
of [Fe/H] = --2.53 with a large dispersion of 0.45 dex. The latter 
suggests that multiple
stellar populations may be present in Coma Berenices. We note that the
Simon \& Geha (2007) metallicity for Coma Berenices is similar to that for
GCs of a comparable luminosity, whereas the Kirby et al. (2008)
metallicity is more metal-poor than the most metal-poor Milky Way
GC and is consistent with 
an extrapolation of the metallicity-luminosity relation to lower
stellar masses.\\

\noindent
$\bullet$ {\bf VUCD7 and F-19}\\
The UCDs VUCD7 in the Virgo cluster 
(Evstigneeva et al. 2007) and F-19 in the Fornax cluster (also known as
UCD3; Mieske et al. 2008) are 
classified as very luminous UCDs with M$_V$ $\sim$ --13.5, 
measured sizes of r$_h$ $\sim$ 90 pc and central velocity
dispersions of $\sigma$ $\sim$ 25 km/s. These values imply masses
of $\sim$ 10$^8$ M$_{\odot}$ and a location within the half-light
zone-of-avoidance. 
However, both of these objects
might be better described as a UCD with an extended ($\sim$ 200
pc) envelope of stars. They may represent transition objects
between nucleated dwarfs and (envelope-free) UCDs. \\

\noindent
$\bullet$ {\bf M59cO}\\
M59cO (also known as SDSS J124155.33+114003.7), located in the
Virgo cluster, was discovered by
Chilingarian \& Mamon (2008). They measured its key properties to
be r$_h$ = 32 pc, $\sigma$ = 48 km/s and M$_V$ $\sim$ --13.5, and
suggested that it is a transition object between UCDs and compact
ellipticals like M32. However its properties place it much closer
to those of UCDs than M32. \\

\noindent
$\bullet$ {\bf NGC 4546 UCD1}\\
Norris \& Kannappan (2010) report the discovery of a UCD
with M$_V$ $\sim$ --13 associated with the nearby S0 galaxy NGC
4546. This UCD is found to have a young age of $\sim$ 3 Gyr and
to be counter-rotating with respect to the stars in NGC 4546
(although, interestingly it co-rotates with the gas around NGC
4546). The high luminosity, young age and retro-grade orbit of
the NGC 4546 UCD would make it a prime candidate for a stellar
system that is {\it not} simply a massive globular cluster of NGC
4546 -- but rather an object that was formed, or accreted, 
in a tidal interaction some 3 Gyrs ago. \\

\noindent
$\bullet$ {\bf Bootes III, Hercules and Ursa Major II}\\
These may be objects in transition between a bound dwarf galaxy
and unbound tidal material.  
In the case of Bootes III, which is on a highly radial orbit, 
Carlin et al. (2009) argue that its
internal kinematics and structure suggest an object in the
process of tidal disruption. It shows evidence for a metallicity
spread in its stars. Hercules is perhaps the most
elongated Milky Way dSph galaxy (apart from the disrupted Sagittarius dwarf
galaxy) with an ellipticity from deep imaging of $\sim$0.65 and a
half-light size of $\sim$170 pc 
(Coleman et al. 2007). We note that its elongation (and other
properties) resemble model RS1-5 of Kroupa (1997). This 
simulation followed the tidal disruption of a dwarf galaxy, in a Milky
Way like halo, that formed without
dark matter (e.g. from condensed gas in a tidal tail). 
Ursa Major II shows signs of ongoing 
tidal interaction (Munoz et al. 2010). Both Hercules and Ursa
Major II reveal evidence for multiple stellar populations (Kirby
et al. 2008).\\

\noindent
$\bullet$ {\bf VCC 2062}\\
Duc et al. (2007) have suggested that VCC 2062, located in the
Virgo cluster, is a tidal dwarf
galaxy formed as the result of an interaction involving NGC 4694
and another galaxy. 
It contains a large quantity of cold gas and exhibits low-level
ongoing star formation, along with evidence of older (0.3 Gyr) stars. 
It has a total luminosity of M$_B$ = --13 and size of a few
kpc. The HI gas reveals a velocity gradient indicative of
rotation. The baryonic (i.e. stellar and cold gas) mass content
accounts for a large fraction of the inferred dynamical mass of
VCC 2062. \\

A summary of how some special case objects match-up to the
different criteria given above is given in Table 1. If the object satisfies
the requirement to be a galaxy it is assigned a $\surd$, a `X' if it fails
and a `?' if it is currently uncertain. As far as we are aware none
of the special case objects hosts a satellite, thus each is
assigned an `X' in that column of Table 1. 
The presence of dark
matter is often controversial and assumes that the measured
velocity dispersion is not dominated by interpolers, binary stars
or tidal heating effects (the latter is questionable for Bootes
III). Under this assumption, two of the
objects have good evidence for a high non-baryonic 
dark matter content. 
Even if we exclude the satellites criterion, none of the objects
listed in Table 1 
satisfies all of the criteria.\\


\begin{table*}
\begin{center}
\caption{Are these stellar systems galaxies?}\label{tableexample}
\begin{tabular}{lccccc}

\hline


Object &  t$_{relax} > H_0^{-1}$ & r$_h >$ 100 pc & Multi-pops &
Dark matter & Satellites\\

\hline
Omega Cen & X & X & $\surd$ & X & X\\
Segue~1 & X & X & $\surd$ & $\surd$$^a$ & X\\
ComBer & $\surd$ & X & ? & $\surd$$^a$ & X\\
VUCD7 & $\surd$ & X & ? & ? & X\\
M59cO & $\surd$ & X & ? & ? & X \\
BooIII &  $\surd$ & $\surd$ & $\surd$ & ? & X\\
VCC2062 & $\surd$ & $\surd$ & $\surd$ & X & X\\
     

\hline
\end{tabular}
\medskip\\
$^a$Presence of dark matter assumes that the observed velocity dispersion
is a valid diagnostic and not due to other effects
(e.g. interpolers, binary stars, tidal heating, non-Newtonian
dynamics etc). 
\end{center}

\end{table*}

\section{Conclusions}

Here we have accepted the popular definition of a galaxy requiring
that it be both gravitationally bound and consist of a system of stars. As
such criteria would include globular (star) clusters, 
additional criteria are required to define a galaxy. 
We suggest that the next best criterion is 
a dynamical one, i.e. that the stars are
collisionless, subject to the general gravitational field of the
system. 
This can be usefully quantified using the two-body relaxation
time.
With these three criteria, globular clusters are effectively
excluded from the definition of a galaxy, as are  
Omega Cen, Segue~1 (and similar objects) and Coma Berenices.
However Ultra Compact Dwarfs
(and perhaps the most massive globular clusters) would be classed as
galaxies. Although this may satisfy some, a fourth criterion
would be required to exclude Ultra Compact Dwarfs. 
We suggest a size-based criterion, e.g. half-light
radius greater than 100 pc. This fourth criterion would exclude the
vast bulk of known Ultra Compact Dwarfs but may still include
extreme objects such as VUCD7 and F-19. 
Bootes III (and similar objects, assuming they are
gravitationally bound) and tidal dwarfs like VCC 2062 would also
be classed as galaxies. 

The combining of criteria above is somewhat subjective and the
opinion of two astronomers. 
The decision of how to define a small planet, and hence the
taxonomic fate
of Pluto, was decided by 424 astronomers present on the
last day of the IAU General Assembly in Prague, held in August 2006.
In order to capture the thoughts of a wider audience about how to define a
galaxy, we invite readers to vote. 
This `collective wisdom' or `crowd-sourcing' 
will be captured in an online poll. The poll allows one to choose
the single best criterion or multiple criteria. 
Results of the poll 
will be reported from time to time at future astronomy conferences. 
The website for anonymous voting is: \\
http://www.surveymonkey.com/s/WLRJMWS\\

\section*{Acknowledgments} 
We thank the following for useful discussions: C. Foster,
A. Graham, D. Koo, E. Ryan-Weber, J. Strader. 

\vspace{10mm}

\noindent{\Large \bf References}
\begin{description}
\item Bailin, J., Harris, W., 2009, ApJ, 695, 1082
\item Barnes, J., Hernquist, L., 1992, Nature, 360, 715
\item Baumgardt, H., Mieske, S., 2008, MNRAS, 391, 942 
\item Baumgardt, H., Cote, P., Hilker, M., Rejkuba, M.,
Mieske, S., Djorgovski, S., Peter, S., 2009, MNRAS, 396, 2051
\item Bekki, K., Couch, W., Drinkwater, M., 2001, ApJ, 552,
105
\item Bekki, K., Freeman, K., 2003, MNRAS, 346, 11
\item Bekki, K., Chiba, M., 2004, A\&A, 417, 437
\item Belokurov, V., et al. 2007, ApJ, 654, 897
\item Belokurov, V., et al. 2009, MNRAS, 397, 1748
\item Binney, J., Treamine, S., 1987, Galactic Dynamics,
Princeton University Press
\item Brada, R., Milgrom, M., 2000, ApJ, 541, 556 
\item Bromm, V., Clarke, C., 2002, ApJ, 566, L1
\item Carlin, J., et al. 2009, ApJ, 702, 9
\item Chiboucas, K., et al. 2010, arXiv:1009.3950
\item Chilingarian, I., Mamon, G., 2008, MNRAS, 385, 83
\item Chilingarian, I., Mieske, M., Hilker, M., Infante,
L., MNRAS, submitted
\item Coleman, M., et al. 2007, ApJ, 668, L43
\item Conroy, C., Loeb, A., Spergel, D., 2010, ApJ, in press
\item Dabringhausen, J., Hilker, M., Kroupa P., 2008, MNRAS,
386, 864 
\item Dabringhausen, J., Kroupa, P., Baumgardt, H., 2009,
MNRAS, 394, 1529
\item Da Rocha, C., Mieske, S., Georgiev, I. Y., Hilker,
M., Ziegler, B. L., 
Mendes de Oliveira, C., arXiv:1009.3010
\item Diemand, J., et al. 2008, Nature, 454, 735
\item Doyle, M., et al. 2005, MNRAS, 361, 34
\item Drinkwater, M., et al.  2000, PASA, 17, 227
\item Drinkwater, M., et al.  2004, PASA, 21, 375
\item Duc, P.-A., et al. 2007, A\&A, 475, 187
\item Evstigneeva, E., Gregg, M., Drinkwater, M.,
Hilker, M., 2007, AJ, 133, 1722
\item Fellhauer, M., Kroupa, P., 2002, MNRAS, 330, 642
\item Forbes, D., 2005, ApJ, 635, 137
\item Forbes D.~A., Lasky P., Graham A.~W., Spitler L.,
2008, MNRAS, 389, 1924
\item Forbes, D., Spitler, L., Graham, A., Foster, C., Hau,
G., Benson, A., 2010, MNRAS, submitted
\item Freeman, K., 1993, ASPC, 48, 608
\item Geha, M., et al. 2009, ApJ, 692, 1464
\item Geha, M., et al. 2010, ApJ, 711, 361
\item Gentile G., Famaey B., Combes F., Kroupa P., Zhao
H.~S., Tiret O., 2007, A\&A, 472, L25
\item Gilmore, G., et al. 2007, ApJ, 663, 948
\item Griffen, B., et al. 2010, MNRAS, 405, 375
\item Hilker, M., et al. 1999, A\&A, 134, 75 
\item Hilker, M., 2009, ariX:0906.0776
\item Hubble, E., 1926, ApJ, 64, 321
\item Kilborn, V., et al. 2005, PASA, 22, 326
\item Kirby, E., Simon, J., Geha, M., Guhathakurta, P.,
Frebel, A., 2008, ApJ, 685, 43L
\item Kroupa, P., 1997, New Astronomy, 2, 139
\item Kroupa, P., 1998, MNRAS, 300, 200
\item Kroupa, P., 2008, In {\it Dynamical Evolution of Dense
Stellar Systems}, IAU Symposium 246, 13
\item Kroupa, P., et al. 2010, A\&A, 523, 32 
\item Lane, R., et al. 2010, MNRAS, 406, 2732
\item Madrid, J., et al. 2010, MNRAS, submitted
\item Mashchenko, S., Sills, A., 2005, ApJ, 619, 258
\item McGaugh, S., Wolf, J., 2010, arXiv:1003.3448
\item Meylan G., Sarajedini A., Jablonka P., Djorgovski
S.~G., Bridges T., Rich R.~M., 2001, AJ, 122, 830
\item Metz M., Kroupa P., Theis C., Hensler G., 
Jerjen H., 2009, ApJ, 697, 269 
\item Mieske, S., Hilker, M., Infante, L., 2004, A\&A, 418,
445
\item Mieske, S., et al. 2008, A\&A, 487, 921
\item Moore, B., 1996, ApJ, 461, L13
\item Munoz, R., Geha, M., Willman, B., 2010, AJ, 140, 138
\item Niederste-Ostholt, M., Belokurov, V., Evans, N.,
Gilmore, G., Wyse, R., Norris, J., 2009, MNRAS, 398, 1771
\item Norris, M., Kannappan, S., arXiv:1009.2489
\item Parmentier, G., 2004, MNRAS, 351, 585
\item Phillipps S., Drinkwater M.~J., Gregg 
M.~D., Jones J.~B., 2001, ApJ, 560, 201
\item Piotto, G., 2009, arXiv:0902.1422
\item Saitoh, T., Koda, J., Okamoto, T., Wada, K., Habe,
A., 2006, ApJ, 640, 22
\item Simon, J., Geha, M., 2007, ApJ, 670, 313
\item Simon, J., et al. 2010, arXiv:1007.4198
\item Soter, S., 2006, AJ, 132, 2513
\item Strader, J., Smith, G., 2008, AJ, 136, 1828
\item Tammann, G., 1994, ESOC, 49, 3
\item Taylor M.~A., Puzia T.~H., Harris G.~L., 
Harris W.~E., Kissler-Patig M., Hilker M., 2010, ApJ, 712, 1191
\item Toloba, E., et al. 2010, A\&A, submitted
\item Verde, L., Oh, S., Jimenez, R., 2002, MNRAS, 336, 541
\item Walsh S.~M., Jerjen H., Willman B., 2007, ApJ, 662, L83 
\item White, S., Rees, M., 1978, MNRAS, 183, 341
\item Willman, B., et al. 2005, ApJ, 626, 85
\end{description}


\end{document}